\documentclass[aps,prd,reprint,eqsecnum,nofootinbib,superscriptaddress,nolongbibliography,floatfix]{revtex4-2}

\usepackage{amsmath}
\usepackage{amssymb}
\usepackage{CJK}
\usepackage{graphicx}
\usepackage{array}
\usepackage{microtype}
\usepackage[colorlinks=true,linkcolor=blue,citecolor=blue,urlcolor=blue,hypertexnames=false]{hyperref}
\hypersetup{pdftitle={Black holes with conical infinity in GCT}}

\newcommand\dx{\mathrm{d}}
\newcommand\rh{r_{\rm h}}
\newcommand\xih{\xi_{\rm h}}
\newcommand\MAF{\mathcal{M}_{\rm AF}}
\newcommand\MH{\mathcal{M}_{\rm H}}
\newcommand\Fcal{\mathcal{F}}

\newcommand\CQG{Classical Quantum Gravity}
\newcommand\JCAP{J. Cosmol. Astropart. Phys.}
\newcommand\LRR{Living Rev. Relativity}

\newcommand\PRD{Phys. Rev. D}

\newcommand\PRL{Phys. Rev. Lett.}

\newcommand\AmJPhys{Am. J. Phys.}

\newcommand\JMathPhysNY{J. Math. Phys. (N.Y.)}

\newcommand\PhilTransRSocA{Phil. Trans. R. Soc. A}

\newcommand\PhysRep{Phys. Rep.}
\newcommand\PhysRev{Phys. Rev.}

\begin{document}

\begin{CJK*}{UTF8}{gbsn}
\title{Black holes with conical infinity in gravitational caloric theory}
\author{S. X. Tian (田树旬)}
\email[]{tshuxun@bnu.edu.cn}
\affiliation{Department of Astronomy, School of Physics and Astronomy, Beijing Normal University, 100875 Beijing, China}
\author{Zong-Hong Zhu (朱宗宏)}
\email[]{zhuzh@bnu.edu.cn}
\affiliation{Department of Astronomy, School of Physics and Astronomy, Beijing Normal University, 100875 Beijing, China}
\affiliation{Institute for Frontiers in Astronomy and Astrophysics, Beijing Normal University, Beijing 102206, China}
\affiliation{School of Physics and Technology, Wuhan University, Wuhan 430072, China}
\date{\today}
\begin{abstract}
  Whether gravitational caloric theory admits spherical black-hole solutions beyond Schwarzschild remains open. A well-defined black-hole solution requires both a regular horizon and a nonsingular exterior extending to infinity. To investigate this question, we cast the vacuum equations for a purely radial caloric field in a general static spherically symmetric metric into a three-dimensional autonomous system, which facilitates the global classification of solutions. We show that the regular asymptotic geometry can be Minkowski or conical, with the latter requiring a specific relation between the model parameters. Within this parameter slice, Schwarzschild is the only asymptotically flat black hole, whereas non-Schwarzschild black holes with conical infinity exist. These black holes can remain arbitrarily close to Schwarzschild from the near-horizon region to the conventional weak-field region (e.g., Solar System scales), yet retain an order-unity solid-angle deficit or excess at infinity. The asymptotic solid angle depends on a single model parameter and is independent of the caloric field amplitude. At fixed locally inferred mass, the caloric amplitude sets the conical transition scale, with weaker fields shifting the transition outward. We derive the local metric corrections and use the resulting periapsis shift to formulate an orbital test of the transition scale.
\end{abstract}
\maketitle
\end{CJK*}

\section{Introduction}
Gravitational caloric theory (GCT) extends general relativity by introducing a caloric vector field~\cite{Tian2026.PRD.114.044042}. Its cosmological applications include early dark energy, a candidate self-tuning mechanism for the old cosmological constant problem, and an early static hot Universe scenario proposed as an alternative to inflation. Beyond cosmology, GCT admits six independent luminal gravitational-wave polarizations around Minkowski spacetime and diverse static spherical geometries~\cite{Tian2026.PRD.114.044042}. However, Schwarzschild was the only black hole identified in the static spherical vacuum sectors examined in that work. Since the numerical search used trial boundary data, it remains unclear whether this result reflects a no-hair restriction~\cite{Israel1967.PhysRev.164.1776,Chrusciel2012.LRR.15.7,Bekenstein1995.PRD.51.R6608} or an incomplete exploration of the solution space. To find black-hole solutions, one must establish regularity of both the metric and caloric field at the horizon, together with a nonsingular exterior extending to infinity. Note that a local near-horizon expansion alone does not establish such a solution.

The structure equations form a system of ordinary differential equations. If they can be cast into a suitable autonomous system, critical points can describe the geometry near spatial infinity or an inner singularity, while the phase portrait shows how these limits are connected. This approach is widely used in cosmological dynamics~\cite{Tian2026.PRD.114.044042,Bahamonde2018.PhysRep.775.1}, but is rarely applied to the structure equations of local gravitational systems~\cite{Mignemi1992.PRD.46.1475,Poletti1994.PRD.50.7260,Ganguly2015.CQG.32.105006,Cruz2017.CQG.34.125014}. In this paper, we explore the dynamical-system approach to studying the global properties of the GCT structure equations. In particular, from a methodological perspective, we find that horizon regularity confines black-hole trajectories to a two-dimensional invariant submanifold of the full three-dimensional phase space. Exploratory integrations, as in Fig.~11 of Ref.~\cite{Tian2026.PRD.114.044042}, can therefore miss black-hole solutions: boundary data chosen at random in the full three-dimensional phase space have zero probability of landing exactly on this two-dimensional surface. Dynamical analysis provides a practical way to locate this submanifold and determine the boundary conditions for a regular horizon.

This paper is organized as follows. Section~\ref{sec:II} presents the relevant aspects of the theory and casts the fluid-free vacuum structure equations into a three-dimensional autonomous system. Section~\ref{sec:III} analyzes the finite critical points and shows how outer regularity constrains the model parameters. Section~\ref{sec:IV} examines the critical points at infinity and constructs the black-hole solutions on the horizon submanifold. Section~\ref{sec:V} analyzes periapsis precession as an observational test and briefly discusses light propagation. Section~\ref{sec:VI} presents conclusions and prospects.

\section{Theory and structure equations}\label{sec:II}
We consider fluid-free vacuum exteriors with $\Lambda=0$ and set $c_i=0$ for $i\notin\{2,4,6\}$ to remove the fluid-dependent terms. The choice $c_2=1$ is motivated by linear stability about empty Minkowski spacetime~\cite{Tian2026.PRD.114.044042}. The remaining dimensionless model parameters, $c_4$ and $c_6$, enter only at nonlinear order about this background and are not fixed by that analysis. With $G_{\mu\nu}$ denoting the Einstein tensor and $S_\mu$ the caloric field, Eqs.~(2.1a) and (2.2) of Ref.~\cite{Tian2026.PRD.114.044042} reduce to
\begin{align}\label{eq:field}
  &G_{\mu\nu} + \nabla_\mu S_\nu + \nabla_\nu S_\mu + c_6S_\mu S_\nu \nonumber\\
    &\quad - g_{\mu\nu} (\nabla_\alpha S^\alpha + c_4S_\alpha S^\alpha) = 0.
\end{align}
The caloric field equation, Eq.~(2.1b) of Ref.~\cite{Tian2026.PRD.114.044042}, then follows from the contracted Bianchi identity. For a static spherically symmetric spacetime, we write the metric in areal-radius coordinates as~\cite{Tolman1939.PhysRev.55.364,Oppenheimer1939.PhysRev.55.374}
\begin{equation}\label{eq:metric}
  \dx s^2 = - e^\nu c^2\dx t^2 + e^\lambda\dx r^2 + r^2\dx\Omega^2,
\end{equation}
where $\nu=\nu(r)$, $\lambda=\lambda(r)$, and $\dx\Omega^2=\dx\theta^2+\sin^2\theta\,\dx\varphi^2$ is the line element on the unit two-sphere. Hereafter, we call $e^{\nu/2}$ the lapse function and $e^\nu$ the temporal metric potential. We take the purely radial configuration $S_\mu=[0,\beta(r),0,0]$, with $[\beta]=\mathrm{length}^{-1}$. This is a special case of Eq.~(5.2) in Ref.~\cite{Tian2026.PRD.114.044042} and satisfies the off-diagonal field equation identically. Spherical symmetry also permits a temporal component, which is not included in the present analysis.

To obtain a first-order system, we introduce three dimensionless variables,
\begin{equation}\label{eq:phase-variables}
  x_1 = e^\lambda,\quad x_2 = r\frac{\dx\nu}{\dx r},\quad x_3 = r\beta,
\end{equation}
which separately describe the radial metric potential, the logarithmic slope of the temporal potential, and the scaled caloric field. We use the logarithmic radius $\ell\equiv\ln(r/r_*)$ as the independent variable, where $r_*>0$ is an arbitrary reference length and $\dx/\dx\ell=r\,\dx/\dx r$. The diagonal components of Eq.~(\ref{eq:field}) are the fluid-free, purely radial specialization of Eqs.~(5.3b)-(5.3d) in Ref.~\cite{Tian2026.PRD.114.044042}. Eliminating $\dx\beta/\dx r$ and $\dx\lambda/\dx r$ from the second-order equation for $\nu$ gives
\begin{subequations}\label{eq:flow}
\begin{align}
  \frac{\dx x_1}{\dx \ell} &= x_1\big[2 - 2x_1 + x_2 - 4x_3 + (c_6 - 2c_4)x_3^2\big], \label{eq:flow-a}\\
  \frac{\dx x_2}{\dx \ell} &= (c_6 - 2c_4)(x_2/2 - 1)x_3^2 - x_1x_2, \label{eq:flow-b}\\
  \frac{\dx x_3}{\dx \ell} &= x_1 - x_2 + 4x_3 - 1 - (c_6 - c_4 + 2)x_3^2  \nonumber\\
    &\quad + (x_2 - x_1)x_3 + (c_6/2 - c_4)x_3^3. \label{eq:flow-c}
\end{align}
\end{subequations}
Equations~(\ref{eq:flow}) form a closed autonomous system in the static region $x_1>0$. Once a trajectory is known, integrating $\dx\nu/\dx\ell=x_2$ reconstructs the temporal potential up to a constant rescaling of time.

Several physical clues and mathematical requirements guide the above construction. There are two motivations for the logarithmic radius $\ell$. First, it parallels the cosmological e-folding number $N=\ln(a/a_0)$, where $a/a_0$ is the normalized scale factor~\cite{Tian2026.PRD.114.044042}. Second, it is suited to the power-law metric asymptotics in Fig.~9 of that reference: a linearized eigenmode $e^{\kappa\ell}$ becomes $(r/r_*)^\kappa$, with $\kappa$ the corresponding eigenvalue. Thus exponential behavior in the flow parameter directly describes a radial power law. Dimensional analysis suggests combinations such as $r\,\dx\nu/\dx r$ and $r\beta$, but does not uniquely determine the variables. As in cosmological applications~\cite{Tian2026.PRD.114.044042,Bahamonde2018.PhysRep.775.1}, the aim is to obtain a closed autonomous system with no explicit $\ell$ dependence and, preferably, polynomial right-hand sides. Earlier black-hole analyses also obtain polynomial systems using dimensionless geometric variables such as $e^{-\lambda/2}$ and $(r/2)e^{-\lambda/2}\,\dx\nu/\dx r$, with an outward radial orientation and a suitable radial parameterization~\cite{Ganguly2015.CQG.32.105006,Cruz2017.CQG.34.125014}. The latter variable combines both metric potentials, whereas $x_1$ and $x_2$ keep their contributions separate, facilitating the analysis of radial power laws and asymptotic geometry. For critical points at infinity, local projective coordinates such as the $z_i$ in Sec.~\ref{sec:IV.A} can regularize divergent directions (see Fig.~6 of Ref.~\cite{Tian2026.PRD.114.044042} for a two-dimensional illustration).

We clarify several conventions and methodological points for the radial analysis. Stability refers to radial flow with increasing $\ell$, rather than to time-dependent perturbations. A physically relevant endpoint need not be a sink or a source, whose eigenvalues (roughly speaking) all have negative or positive real parts, respectively. At a saddle, solutions of the full nonlinear system can lie on its stable or unstable invariant manifold and approach the point as $r$ increases or decreases, respectively. For example, the saddle $F_1$ below describes the asymptotically flat end of Schwarzschild along its stable manifold and the approximately Minkowskian core of the conically dressed Minkowski metric along its unstable manifold. The focus is thus on trajectories selected by spatial boundary conditions, which differs from the common cosmological search for attractors reached from generic initial conditions~\cite{Tian2026.PRD.114.044042,Copeland1998.PRD.57.4686}. Hereafter, a \textit{finite critical point} is one at which all $x_i$ remain finite, while a \textit{critical point at infinity} represents a direction in which at least one $x_i$ diverges. These terms refer to phase space, not to the physical radius. At phase-space infinity, reconstructing the radius distinguishes between spatial infinity and a finite-radius boundary. For the black holes with conical infinity below, a finite critical point describes spatial infinity, while a critical point at phase-space infinity describes the horizon at the inner boundary of the static exterior. In either case, regularity must be checked for both the metric and caloric field.

\section{Finite critical points and asymptotic geometry}\label{sec:III}
We first examine the finite critical points to determine the possible outer geometries and the associated constraints on the model parameters. For the outer geometry, we require $e^\lambda$ and $e^\nu$ to approach finite positive constants at spatial infinity. A finite critical point can also describe a regular center when approached inward. Limits with $x_1\to0$ require a separate analysis and are not part of the classification of flat and conical infinity below.

At a finite critical point with $x_1>0$, integrating $\dx\nu/\dx\ell=x_2$ gives $e^\nu\propto r^{x_2}$. A finite nonzero asymptotic lapse therefore requires $x_2=0$, and Eq.~(\ref{eq:flow-b}) reduces to $(c_6-2c_4)x_3^2=0$. If $c_6\ne2c_4$, then $x_3=0$, and the remaining equations give $F_1=(1,0,0)$. This point exists for all real $c_4$ and $c_6$. Its eigenvalues $(-1,-1,2)$ identify a saddle with a two-dimensional stable manifold and a one-dimensional unstable manifold. The stable manifold describes asymptotically flat ends as $\ell\to+\infty$, while the unstable manifold describes regular centers as $\ell\to-\infty$, or $r\to0$. Table~\ref{tab:critical} summarizes these two interpretations, and Fig.~\ref{fig:manifolds} shows the stable manifold in blue for a representative parameter choice.

\begin{table*}[t]
  \everymath{\displaystyle}
  \caption{Finite critical points with $x_1>0$ and critical points at infinity in the chart $(z_1,z_2,z_3)$, for $c_6=2c_4$. Eigenvalues refer to increasing $\ell$ or $\sigma$, as indicated. The last column gives representative physical radial limits, without classifying every parameter range and stable or unstable manifold. Inner and outer denote lower and upper areal radii within a static region. At $I_1$, $I_2$, $I_3$, and $I_4$, the eigenvalues normal to $z_1=0$ are $2$, $1$, $2c_4/(c_4-2)$, and $1$, respectively. Positive (negative) normal eigenvalues allow physical approaches only inward (outward). Tangential stable directions need not be accessible from $z_1>0$. The point $F_2$ is physical for $c_4<0$ or $c_4>2$. The points $I_3$ and $I_4$ require $c_4\ne2$ and coincide at $c_4=-2$. Manifold labels assume nonzero eigenvalues. For $c_4=0$, the coordinate singularity at infinity associated with $I_3$ has vanishing curvature and is neither asymptotically flat nor conical.}
  \label{tab:critical}
  \begin{tabular*}{\textwidth}{@{\extracolsep{\fill}}lccl@{}}
  \hline\hline
  Critical point & Coordinates & Eigenvalues & Physical radial limit\\
  \hline
  Finite & $(x_1,x_2,x_3)$ & (radial flow in $\ell$) & \\[1pt]
  $F_1$ & $(1,0,0)$ & $-1,-1,2$ & Flat infinity (stable manifold);\\
  & & & regular center (unstable manifold)\\[1pt]
  \cline{2-4}
  \noalign{\vskip 2pt}
  $F_2$ & $\bigl(1-\frac{2}{c_4},0,\frac{1}{c_4}\bigr)$ & $\dfrac{2-c_4}{c_4},\ \dfrac{2-c_4\pm\sqrt{4+12c_4-7c_4^2}}{2c_4}$ & Conical infinity (outward sink)\\
  \noalign{\vskip 2pt}
  \hline
  Infinity & $(z_1,z_2,z_3)$ & (radial flow in $\sigma$) & \\[1pt]
  $I_1$ & $(0,0,0)$ & $1,1,2$ & Inner throat (outward source)\\[1pt]
  \cline{2-4}
  \noalign{\vskip 1pt}
  $I_2$ & $(0,1,0)$ & $-1,1,1$ & Inner horizon on $\MH$ (unstable manifold)\\[1pt]
  \cline{2-4}
  \noalign{\vskip 2pt}
  $I_3$ & $\bigl(0,0,\frac{1}{c_4-2}\bigr)$ & $-1,\ \dfrac{2c_4}{c_4-2},\ \dfrac{c_4+2}{c_4-2}$ & Inner field singularity: $c_4<0$ or $c_4>2$;\\
  & & & outer field singularity: $0<c_4<2$;\\
  & & & coordinate singularity at infinity: $c_4=0$\\[1pt]
  \cline{2-4}
  \noalign{\vskip 2pt}
  $I_4$ & $\bigl(0,\frac{c_4+2}{c_4-2},\frac{1}{c_4-2}\bigr)$ & $-1,\ 1,\ -\dfrac{c_4+2}{c_4-2}$ & Inner singularity (unstable manifold)\\
  \noalign{\vskip 2pt}
  \hline\hline
  \end{tabular*}
\end{table*}

A nonzero limiting value of the scaled caloric field $x_3=r\beta$ is compatible with the above asymptotic conditions only if
\begin{equation}\label{eq:conical-slice}
  c_6 = 2c_4.
\end{equation}
Nonzero fields with $r\beta\to0$ can instead approach $F_1$ without this restriction. Equation~(\ref{eq:conical-slice}) is independent of the Minkowski stability argument for $c_2=1$ and is imposed throughout the remainder of the paper. With $x_2=0$, the critical-point equations become $x_1=1-2x_3$ and $x_3(1-c_4x_3)=0$. Besides $F_1$, they admit $F_2=(1-2/c_4,0,1/c_4)$ for $c_4\ne0$. The condition $x_1>0$ restricts this second point to $c_4<0$ or $c_4>2$. Figure~\ref{fig:manifolds} Left illustrates the case $c_4=4$, for which $F_1=(1,0,0)$ and $F_2=(1/2,0,1/4)$ describe distinct outer geometries.

Whether nearby solutions approach the geometry at $F_2$ is determined by its radial stability. One eigenvalue is $2/c_4-1$, while the other two have sum $2/c_4-1$ and product $2-4/c_4$. All three therefore have negative real parts in both physical parameter ranges. Thus $F_2$ is a hyperbolic sink with a three-dimensional stable manifold. For $c_4=4$, the complex pair $-1/4\pm i\sqrt{15}/4$ in Table~\ref{tab:critical} explains the damped radial oscillations in Fig.~\ref{fig:bh-profile}.

Along a trajectory approaching $F_2$, exponential decay of $x_2$ in $\ell$ makes $\nu$ tend to a finite constant. After a constant rescaling of time, the asymptotic metric is
\begin{equation}\label{eq:cone}
  \lim_{r\to\infty}\dx s^2 = -c^2\dx t^2 + \left(1-\frac{2}{c_4}\right)\dx r^2 + r^2\dx\Omega^2.
\end{equation}
On a static slice, the proper radial distance from a fixed sphere $r=r_0$ grows as $\sqrt{1-2/c_4}\,r$. The limiting ratio of the sphere's area to the square of this distance defines the solid angle at infinity,
\begin{equation}\label{eq:solid-angle}
  \lim_{r\to\infty}\frac{4\pi r^2}{\left[\int_{r_0}^{r}e^{\lambda(\tilde r)/2}\,\dx\tilde r\right]^2} = \frac{4\pi c_4}{c_4-2}.
\end{equation}
The result is independent of $r_0$ and gives a solid-angle deficit for $c_4<0$ and an excess for $c_4>2$. Such an angular deformation is compatible with vanishing curvature at infinity~\cite{Nucamendi1997.CQG.14.1309}, as in the asymptotic geometries of global monopoles and black holes in bumblebee and scalar-tensor gravity~\cite{Barriola1989.PRL.63.341,Casana2018.PRD.97.104001,Chagoya2018.JCAP.08.006}. Here, $S_\mu\approx[0,1/(c_4r),0,0]$ and $S_\mu S^\mu\approx[c_4(c_4-2)r^2]^{-1}$. Although the caloric field vanishes at infinity, its scaled value approaches the nonzero limit $r\beta\to1/c_4$, which supports the conical geometry. The leading curvature invariants are
\begin{equation}\label{eq:cone-curvature}
  R \approx \frac{4}{(2-c_4)r^2},\quad K \approx \frac{16}{(c_4-2)^2r^4},
\end{equation}
where $R$ is the Ricci scalar and $K\equiv R_{\mu\nu\rho\sigma}R^{\mu\nu\rho\sigma}$ is the Kretschmann scalar. Both vanish at infinity despite the finite angular deformation. The equilibrium solution at $F_2$ is the pure conical metric: the right-hand side of Eq.~(\ref{eq:cone}) holds exactly for every $r>0$ and has a singular center at $r=0$. The nonconstant solutions below approach $F_2$ only outward, so their inner geometries need not share this singularity.

The radial flow also determines how the temporal potential approaches its asymptotic value. Since $F_2$ is a hyperbolic sink, $x_1-(1-2/c_4)$ is integrable in $\ell$. For a horizon branch approaching $F_2$, integrating $\dx\ln x_2/\dx\ln r=-x_1$ and $\dx\nu/\dx\ln r=x_2$ gives
\begin{equation}\label{eq:conical-falloff}
  1-e^\nu \approx \frac{c_4x_2}{c_4-2} \propto \left(\frac{\rh}{r}\right)^{1-2/c_4}.
\end{equation}
Here $\rh$ is the horizon radius, and the positive proportionality factor depends on the branch. The exponent $1-2/c_4$ is positive in both physical parameter ranges, making the normalization $e^\nu\to1$ well defined. This falloff generally differs from the Schwarzschild behavior. Consequently, $\rh$ determines the horizon area but does not by itself define a mass through $\rh=2GM/c^2$. A global mass definition must account for the nonflat asymptotic background~\cite{Nucamendi1997.CQG.14.1309}.

\section{Critical points at infinity and inner geometry}\label{sec:IV}
To construct a black-hole solution, the outer geometry must be connected to a regular horizon. We first examine the critical points at infinity and distinguish horizons from throats and singular radial limits. We then identify the horizon submanifold, establish Schwarzschild uniqueness under asymptotic flatness, and construct black holes with conical infinity. The last part of the section explains their local Schwarzschild limit and the scale of the conical transition.

\subsection{Critical points at infinity}\label{sec:IV.A}
A divergence of $x_i$ need not indicate a physical singularity, since areal-radius coordinates can fail at a regular throat or horizon. To examine directions in which $x_1\to\infty$ while $x_2/x_1$ and $x_3/x_1$ remain finite,\footnote{For the metric~(\ref{eq:metric}), contracting the curvature components in Eqs.~(5)-(7) of Ref.~\cite{Zaslavskii2007.PRD.76.024015} and using Eqs.~(\ref{eq:flow-a}) and~(\ref{eq:flow-b}) with $c_6=2c_4$ gives $r^4K=\{4x_2^2(x_3-1)^2+2x_2^2+4(x_1-1)^2+2[x_2-4x_3+2(1-x_1)]^2\}/x_1^2$. Every term is nonnegative. Thus finite $K$ at finite positive $r$ bounds $x_2/x_1$ and $(x_1-1)/x_1$, and the last square then bounds $x_3/x_1$. The present chart therefore covers every regular horizon at such a radius. A complementary chart for $x_3\to+\infty$ can instead be defined by $(z_1,z_2,z_3)=(x_1,x_2,1)/x_3$. For $c_4=4$, its regularized flow has an additional critical point $(0,0,0)$, which is an outward source. Physical inward approaches reach a finite positive radius with $x_1\propto x_3^{2/3}$ and finite $x_2$, implying $x_3/x_1\to\infty$ and $K\to\infty$. The remaining complementary directions likewise describe singular endpoints or are inaccessible from the static region, and supply no additional regular horizon.} we introduce the projective coordinates~\cite{Tian2026.PRD.114.044042,Perko2001.book}
\begin{equation}\label{eq:projective-variables}
  z_1 = \frac{1}{x_1},\quad z_2 = \frac{x_2}{x_1},\quad z_3 = \frac{x_3}{x_1},
\end{equation}
together with the rescaled radial parameter $\sigma$ defined by $\dx\ell/\dx\sigma=z_1$. This rescaling removes the $1/z_1$ factors in the projected equations, allowing a regular extension to $z_1=0$. Since $z_1>0$ in the static region, increasing $\sigma$ corresponds to increasing radius. With $c_6=2c_4$, Eq.~(\ref{eq:flow}) becomes
\begin{subequations}\label{eq:zflow}
\begin{align}
  \frac{\dx z_1}{\dx\sigma} &= z_1(2 - 2z_1 - z_2 + 4z_3),\\
  \frac{\dx z_2}{\dx\sigma} &= z_2(1 - 2z_1 - z_2 + 4z_3),\\
  \frac{\dx z_3}{\dx\sigma} &= z_1 + z_3 - z_1(z_1 + z_2 - 2z_3) + (2 - c_4)z_3^2.
\end{align}
\end{subequations}

\begin{figure*}[t]
  \centering
  \includegraphics[width=0.98\textwidth]{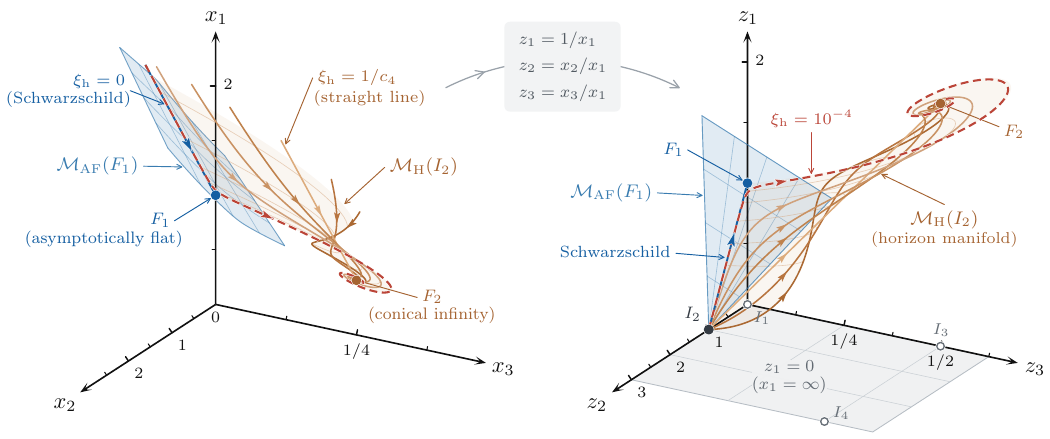}
  \caption{Representative black-hole solutions in phase space for $c_4=4$ and $c_6=2c_4$. The same eight horizon trajectories are shown in the $x_i$ (Left) and $z_i$ (Right) coordinates. The brown mesh samples the horizon submanifold $\MH(I_2)$. The blue mesh (Left) lies on the asymptotically flat manifold $\MAF(F_1)$, whereas the blue triangle (Right) is a schematic representation of this manifold near $I_2$. All trajectories are plotted at their computed coordinates. The solid blue Schwarzschild trajectory ($\xih=0$) connects $I_2$ to $F_1$ along the intersection of the two manifolds. The red dashed trajectory ($\xih=10^{-4}$) passes near $F_1$ before approaching $F_2$, where the asymptotic solid angle is $8\pi$. Its radial metric profile appears in Fig.~\ref{fig:bh-profile}. The pale brown straight line is the exact solution~(\ref{eq:exact}) with $\xih=1/c_4=1/4$. The other solid brown trajectories have $\xih=1/64$, $1/16$, $1/8$, $1/2$, and $1$. Arrows indicate increasing radius, and the gray plane is the boundary $z_1=0$. Mesh edges only delimit the displayed portions, and apparent crossings need not be intersections in three dimensions. Taking the local Schwarzschild mass to zero at fixed transition scale gives the horizonless $F_1\to F_2$ trajectory of the conically dressed Minkowski metric, with an approximately Minkowskian core and conical infinity.}
  \label{fig:manifolds}
\end{figure*}

On the invariant boundary $z_1=0$, the last two equations reduce to $\dx z_2/\dx\sigma=z_2(1-z_2+4z_3)$ and $\dx z_3/\dx\sigma=z_3[1+(2-c_4)z_3]$. Their simultaneous zeros give the four critical points $I_1$, $I_2$, $I_3$, and $I_4$, whose coordinates and full three-dimensional eigenvalues are listed in Table~\ref{tab:critical}. These points lie on the gray boundary plane in Fig.~\ref{fig:manifolds} Right, while $F_1$ and $F_2$ lie in the physical region $z_1>0$. Thus the projective chart places the candidate inner boundaries and the outer geometries of Sec.~\ref{sec:III} in the same phase portrait.

The signs of the eigenvalues must be interpreted together with the directions accessible from the physical region. A positive eigenvalue normal to $z_1=0$ permits an approach from $z_1>0$ only inward, whereas a negative one permits it only outward.\footnote{Equation~(\ref{eq:zflow}) implies that $\dx\ln z_1/\dx\sigma$ tends to the normal eigenvalue. A positive value excludes $z_1\to0$ as $\sigma\to+\infty$, while a negative value excludes it as $\sigma\to-\infty$. Along an allowed approach with a nonzero normal eigenvalue, $z_1$ decays exponentially. Integrating $\dx\ln r/\dx\sigma=z_1$ then gives a finite nonzero limiting radius.} Stable directions tangent to the boundary do not by themselves identify a physical outer endpoint. The radius and temporal potential follow by integrating $\dx\ln r/\dx\sigma=z_1$ and $\dx\nu/\dx\sigma=z_2$. We must also check the caloric field, whose squared norm is
\begin{equation}\label{eq:infinity-field-norm}
  S_\mu S^\mu = \frac{x_3^2}{x_1r^2} = \frac{z_3^2}{z_1r^2}.
\end{equation}
If this invariant diverges, the caloric field is singular even when the spacetime curvature remains finite.

The point $I_1$ has three positive eigenvalues, so an open set of physical trajectories approaches it inward. Reconstructing the geometry gives a smooth throat with a nonzero lapse, at which both the metric and caloric field are regular in a proper-distance coordinate~\cite{Ellis1973.JMathPhysNY.14.104,Morris1988.AmJPhys.56.395}. At $I_3$, the normal eigenvalue is positive for $c_4<0$ or $c_4>2$, allowing a physical approach only inward. For $0<c_4<2$, it is negative, and $I_3$ is an outward sink at a finite upper radius. At $I_4$, the normal eigenvalue is always $1$, so physical approaches are again inward. At the finite-radius limits associated with $I_3$ and $I_4$, nonzero $z_3$ makes the caloric-field norm diverge. The geometry at $I_4$ also has a curvature singularity, except at the merger $c_4=-2$.\footnote{The exceptional cases in Table~\ref{tab:critical} require nonlinear analysis. For $c_4=0$, the normal eigenvalue at $I_3$ vanishes. Reconstructing the geometry gives a divergent radial metric component at infinity, vanishing curvature, and a finite nonzero caloric-field norm. Following Ref.~\cite{Tian2026.PRD.114.044042}, we call this a coordinate singularity at infinity. It is neither asymptotically flat nor conical. At $c_4=-2$, the merged point $I_3=I_4$ has a zero tangential eigenvalue but a positive normal one, so its physical branches still describe inner field singularities. At $c_4=2$, both points leave the finite part of this chart.} For $c_4=4$, all three points describe inner radial limits, as shown in Fig.~\ref{fig:manifolds}, and the stable directions at $I_3$ and $I_4$ lie entirely on $z_1=0$.

The remaining point, $I_2=(0,1,0)$, has eigenvalues $(-1,1,1)$. Its positive normal eigenvalue allows only an inward approach, along which $\dx\nu/\dx\sigma\to1$ and the lapse vanishes. This suggests a black-hole horizon. To establish this interpretation, we next identify the relevant invariant manifold and check regularity. The same analysis shows that every smooth finite-area horizon in this sector corresponds to $I_2$.

\subsection{The horizon submanifold}\label{sec:IV.B}
At $I_2$, the eigendirection with eigenvalue $-1$ is tangent to the exact invariant curve $z_1=z_3=0$, on which $\dx z_2/\dx\sigma=z_2(1-z_2)$. This curve is the local stable manifold and lies entirely on the phase-space boundary. Physical trajectories can therefore approach $I_2$ inward only along its two-dimensional unstable manifold~\cite{Perko2001.book}. We denote this surface by $\MH$ and call it the \textit{horizon submanifold}. With stability defined for increasing $\sigma$, physical trajectories leave $I_2$ outward along $\MH$, while the negative eigenvalue describes local transverse attraction toward this surface. Figure~\ref{fig:manifolds} shows $\MH(I_2)$ in brown. The blue surface through $F_1$ represents the separate condition of asymptotic flatness, examined in Sec.~\ref{sec:IV.C}.

Along these trajectories, $z_1\propto e^\sigma$ as $\sigma\to-\infty$, so integrating $\dx\ln r/\dx\sigma=z_1$ gives a finite horizon radius $\rh>0$. Meanwhile, $x_2/x_1\to1$ and $x_3=z_3/z_1$ tends to a free finite limit, denoted by $\xih\equiv\rh\beta(\rh)$. This is the dimensionless caloric amplitude at the horizon. The trajectories near $I_2$ satisfy $z_2=1+(2\xih-1)z_1+\mathcal{O}(z_1^2)$ and $z_3=\xih z_1+\mathcal{O}(z_1^2)$. Thus the distinct trajectories emerging from $I_2$ in Fig.~\ref{fig:manifolds} carry different horizon fields. At fixed $c_4$, the pair $(\rh,\xih)$ specifies a local regular solution up to time normalization.

To check regularity at the horizon, it is useful to replace the divergent metric variables with $f_1(r)\equiv e^{-\lambda}=x_1^{-1}$ and $f_2(r)\equiv e^{\nu+\lambda}$. With $f_{2{\rm h}}\equiv f_2(\rh)>0$ and $\varepsilon\equiv(r-\rh)/\rh$, the local expansion is
\begin{subequations}\label{eq:horizon-series}
\begin{align}
  \beta &= \frac{\xih}{\rh}[1 - c_4\xih\varepsilon + \mathcal{O}(\varepsilon^2)], \label{eq:horizon-beta}\\
  f_1 &= \varepsilon + (\xih - 1)\varepsilon^2 + \mathcal{O}(\varepsilon^3), \label{eq:horizon-f1}\\
  \frac{f_2}{f_{2{\rm h}}} &= 1 - \frac{2}{3}\xih(1 - c_4\xih)\varepsilon^2 + \mathcal{O}(\varepsilon^3). \label{eq:horizon-f2}
\end{align}
\end{subequations}
Introducing the ingoing coordinate $v$ through $c\,\dx v=c\,\dx t+\dx r/(f_1\sqrt{f_2})$ removes the pole in $g_{rr}=1/f_1$ and gives
\begin{equation}\label{eq:ingoing}
  \dx s^2 = - f_1f_2c^2\dx v^2 + 2c\sqrt{f_2}\,\dx v\dx r + r^2\dx\Omega^2.
\end{equation}
The determinant of the $(v,r)$ block is $-c^2f_2$, which remains nonzero at $\rh$. The expansions above therefore give a smooth metric and a smooth caloric covector $S_\mu=[0,\beta(r),0,0]$ in this chart. Outgoing radial null rays obey $\dx r/\dx v=cf_1\sqrt{f_2}/2$. Immediately inside the horizon, $f_1<0$, so even outgoing rays move toward smaller areal radius. This identifies a locally trapped black-hole interior.

To determine whether a regular horizon can lie outside $\MH$, we use the angular component of the trace-reversed field equation. With $c_6=2c_4$, its form in areal-radius coordinates is
\begin{equation}\label{eq:angular}
  \frac{r}{2}\left(\frac{\dx f_1}{\dx r} + f_1\frac{\dx\nu}{\dx r}\right) = 1 - f_1 + 2rf_1\beta.
\end{equation}
Near a smooth finite-area horizon, we introduce a regular transverse Gaussian null coordinate $u$, with the horizon at $u=0$. This avoids assuming that $r$ is already a valid coordinate there. The metric can be written as
\begin{equation}\nonumber
  \dx s^2 = - f_3(u)c^2\dx v^2 + 2c\,\dx v\dx u + r(u)^2\dx\Omega^2.
\end{equation}
For $f_3(0)=0$ and a smooth caloric covector, the angular field equation gives $\rh[(\dx f_3/\dx u)(\dx r/\dx u)]_{u=0}=1$. Both derivatives must be nonzero, so the horizon is nondegenerate and $r$ is a valid local coordinate. Writing $e^\nu=f_1f_2$ with smooth $f_2>0$, Eq.~(\ref{eq:angular}) then fixes $\rh\,\dx f_1/\dx r|_{r=\rh}=1$. Hence $(z_1,z_2,z_3)\to(0,1,0)$, and every such horizon belongs to $\MH$. The positive slope places the static region at larger $r$, excluding extremal and outer cosmological horizons in this sector. A connected static exterior can therefore have at most one regular horizon.

For $\xih\ne0$, the caloric covector is nonzero and null on the future horizon, $S_\mu=[0,\xih/\rh,0,0]$, and $S^\mu$ is tangent to the horizon generator. Its squared norm, $S_\mu S^\mu=f_1\beta^2\approx\xih^2(r-\rh)/\rh^3$, changes sign across the horizon, so the vector is spacelike outside and timelike inside. The horizon curvature $K_{\rm h}=4[(\xih-1)^2+2]/\rh^4$ is finite and depends on $\xih$, although the slope of $f_1$ is universal. These local results hold for all real $c_4$ and $\xih$ and establish the horizon interpretation of $I_2$ in Table~\ref{tab:critical}. The outer geometry is determined by the subsequent trajectory on $\MH$.

\subsection{Schwarzschild uniqueness}\label{sec:IV.C}
An asymptotically flat black-hole exterior must lie on both the horizon submanifold $\MH$ and the asymptotically flat manifold $\MAF(F_1)$ in Fig.~\ref{fig:manifolds}. To determine their intersection, we construct the local stable manifold of $F_1$ by solving the stable-manifold invariance equation order by order~\cite{Tian2026.PRD.114.044042,Perko2001.book}. Such an expansion need not terminate in general, but for Eq.~(\ref{eq:flow}) with $c_6=2c_4$, it terminates exactly at quadratic order, yielding the invariant surface $\Fcal=0$, where
\begin{equation}\label{eq:invariant-function}
  \Fcal \equiv x_1 - x_2 - 1 + (x_2 + 4)x_3 - (c_4 + 2)x_3^2.
\end{equation}
The field equations give $\Fcal/(x_1r^2)=\nabla_\mu S^\mu-2S_\mu S^\mu$, so the condition $\Fcal=0$ has a covariant meaning. The invariance of this surface is verified directly from Eq.~(\ref{eq:flow}), which gives
\begin{equation}\label{eq:normal}
  \frac{\dx}{\dx \ell}\left(\frac{\Fcal}{x_1}\right) = 2(1 - c_4x_3)\frac{\Fcal}{x_1}.
\end{equation}
An asymptotically flat exterior approaches $F_1$, where $\Fcal/x_1\to0$. However, near $F_1$, Eq.~(\ref{eq:normal}) would make any nonzero $\Fcal/x_1$ grow outward at an asymptotic logarithmic rate of $2$. Asymptotic flatness therefore requires $\Fcal$ to vanish along the entire trajectory. The asymptotically flat manifold $\MAF$ is the subset of this invariant surface whose trajectories approach $F_1$. Note that $F_1$ may not be a global attractor on this whole surface. The blue mesh in Fig.~\ref{fig:manifolds} Left samples this asymptotically flat branch. Its connection to the horizon point along the Schwarzschild trajectory is illustrated schematically in Fig.~\ref{fig:manifolds} Right.

On $\MAF$, the flow equations give $\dx(x_3/x_2)/\dx\ell=0$. Along a regular horizon branch, $x_2\to\infty$ while $x_3$ remains finite, so this constant ratio must vanish. Thus $x_3=0$ throughout the exterior, with $S_\mu=0$ and $x_1=1+x_2$. After normalizing time, the metric is $e^\nu=e^{-\lambda}=1-\rh/r$. Horizon regularity and asymptotic flatness therefore select only Schwarzschild in the static purely radial caloric sector with $c_6=2c_4$, for every real $c_4$. The intersection of $\MAF$ and $\MH$ is the Schwarzschild trajectory, shown as the solid blue segment $\{z_3=0,\ z_1+z_2=1\}$ from $I_2$ to $F_1$ in Fig.~\ref{fig:manifolds}. A regular black-hole exterior with $\xih\ne0$ therefore cannot be asymptotically flat. This is different from the situation in many modified gravity theories, such as scalar-Gauss-Bonnet gravity, which can support asymptotically flat black holes with scalar hair~\cite{Sotiriou2014.PRL.112.251102}.

Note that the uniqueness theorem proved here assumes $c_6=2c_4$. We do not extend the black-hole analysis to independent $c_4$ and $c_6$ in this paper, but focus on the black holes with conical infinity constructed below.

\subsection{Black holes with conical infinity}
We now construct black-hole exteriors connecting the regular horizon at $I_2$ to conical infinity at $F_2$. Reducing the system to a single second-order equation gives an exact solution and a convenient formulation for numerical integration. Varying the horizon caloric amplitude then reveals how a solution can remain close to Schwarzschild over an extended region before developing conical geometry. We also compare this behavior with other conical exteriors.

\subsubsection{Existence and construction}
A black hole with conical infinity corresponds to an $I_2\to F_2$ trajectory on the brown surface in Fig.~\ref{fig:manifolds}. The quantity $\Fcal/x_1$ tends to $\xih$ at the horizon and equals $1/c_4$ at $F_2$. Equation~(\ref{eq:normal}) preserves its sign, so $\xih$ and $c_4$ must have the same sign. Together with $x_1\to1-2/c_4>0$, this gives the necessary conditions
\begin{equation}\nonumber
  \{c_4 < 0,\ \xih < 0\}\quad\text{or}\quad\{c_4 > 2,\ \xih > 0\}.
\end{equation}
To determine whether these conditions yield a global black-hole exterior, we integrate outward from regular horizon data at fixed $c_4$ and varying $\xih$, checking both exterior regularity and convergence to the conical critical point.

For this integration, we introduce $\mu\equiv1+x_2-x_1$, which cancels the common horizon divergence of $x_1$ and $x_2$. The expansion in Sec.~\ref{sec:IV.B} gives $\mu=1+(z_2-1)/z_1\to2\xih$ near $I_2$. The inverse redshift gradient is a useful independent variable. Along a regular horizon branch, $x_2>0$ and $\dx x_2/\dx\ell=-x_1x_2<0$, so $\eta\equiv1/x_2=1/(r\,\dx\nu/\dx r)$ increases monotonically from $\eta=0$ at the horizon. This variable was also used in Eq.~(5.8) of Ref.~\cite{Tian2026.PRD.114.044042}. The first two flow equations give $\dx\eta/\dx\ell=1+\eta(1-\mu)=\eta x_1>0$ and $\eta\,\dx\mu/\dx\eta=4x_3-2\mu$. Eliminating $x_3$ reduces the system to
\begin{align}\label{eq:master}
  &\eta[1 + \eta(1 - \mu)]\frac{\dx^2\mu}{\dx \eta^2} + [3 + (c_4 - 2)\eta\mu]\frac{\dx\mu}{\dx \eta} \nonumber\\
  &\quad{}+ \frac{c_4+2}{4}\eta^2 \bigl(\frac{\dx\mu}{\dx \eta}\bigr)^2 + \mu(c_4\mu - 2) = 0.
\end{align}
The regular horizon expansion supplies the boundary data
\begin{equation}\label{eq:horizondata}
  \left.\mu\right|_{\eta=0} = 2\xih,\quad \left.\frac{\dx\mu}{\dx \eta}\right|_{\eta=0} = \frac{4}{3}\xih(1 - c_4\xih).
\end{equation}
The second condition follows from the first by taking the regular $\eta\to0$ limit of Eq.~(\ref{eq:master}). Although this equation is second order, its highest-derivative coefficient vanishes at the horizon. Linearization about a regular branch admits an additional mode with leading behavior $\eta^{-2}$, which is excluded by horizon regularity. Hence $\xih$ is the only free dimensionless horizon datum at fixed $c_4$, and the admissible trajectories lie on $\MH$. This explains why the horizon conditions are essential: a trajectory attracted to $F_2$ need not originate at a regular horizon. A global black-hole solution requires both Eq.~(\ref{eq:horizondata}) and a nonsingular continuation to infinity.

Once $\mu(\eta)$ is known, the original variables are recovered algebraically as $x_1=1+\eta^{-1}-\mu$, $x_2=\eta^{-1}$, and $x_3=(2\mu+\eta\,\dx\mu/\dx\eta)/4$. The radius and metric normalization follow from
\begin{equation}\nonumber
  \frac{\dx\ln r}{\dx\eta} = \frac{1}{1+\eta(1-\mu)},\quad \frac{\dx\ln f_2}{\dx\eta} = - \frac{1}{x_1}\frac{\dx\mu}{\dx\eta}.
\end{equation}
Integrating from $\{r=\rh,\ f_2=f_{2{\rm h}}\}$ at $\eta=0$ gives $e^\nu=f_2/x_1$, $e^\lambda=x_1$, and $\beta=x_3/r$. Numerically, we start at small positive $\eta$ using the regular horizon series. The constant $f_{2{\rm h}}$ sets the time normalization. For a branch approaching $F_2$, choosing $e^\nu\to1$ gives $f_2\to1-2/c_4$. Neither $\mu$ nor $\beta$ is assigned an outer boundary value.

The constant solution $\mu=2/c_4$ of Eq.~(\ref{eq:master}) gives an exact member with $\xih=1/c_4$. After normalizing $e^\nu\to1$, its metric functions are
\begin{equation}\label{eq:exact}
  e^\nu = \left(1-\frac{2}{c_4}\right)e^{-\lambda} = 1-\left(\frac{\rh}{r}\right)^{1-2/c_4},
\end{equation}
with $S_\mu=[0,1/(c_4r),0,0]$ and $c_4<0$ or $c_4>2$. For $c_4=4$, the phase-space relations $x_1=1/2+x_2$ and $x_3=1/4$ give the pale brown straight line in Fig.~\ref{fig:manifolds}. Its projective image is also straight, with $z_2=1-z_1/2$ and $z_3=z_1/4$. Other members are obtained by varying $\xih$ at fixed $c_4$ and $\rh$. Among regular horizon branches in these parameter ranges, only Schwarzschild and Eq.~(\ref{eq:exact}) have constant $f_2$, so the generic metric potentials cannot be made reciprocal by rescaling time. For the exact solution, $\xih=1/c_4$ ties the horizon field to the asymptotic angle. The general family instead has an independent horizon caloric amplitude, which allows a local Schwarzschild limit at fixed asymptotic angle.

\begin{figure}[tb]
  \centering
  \includegraphics[width=0.98\columnwidth]{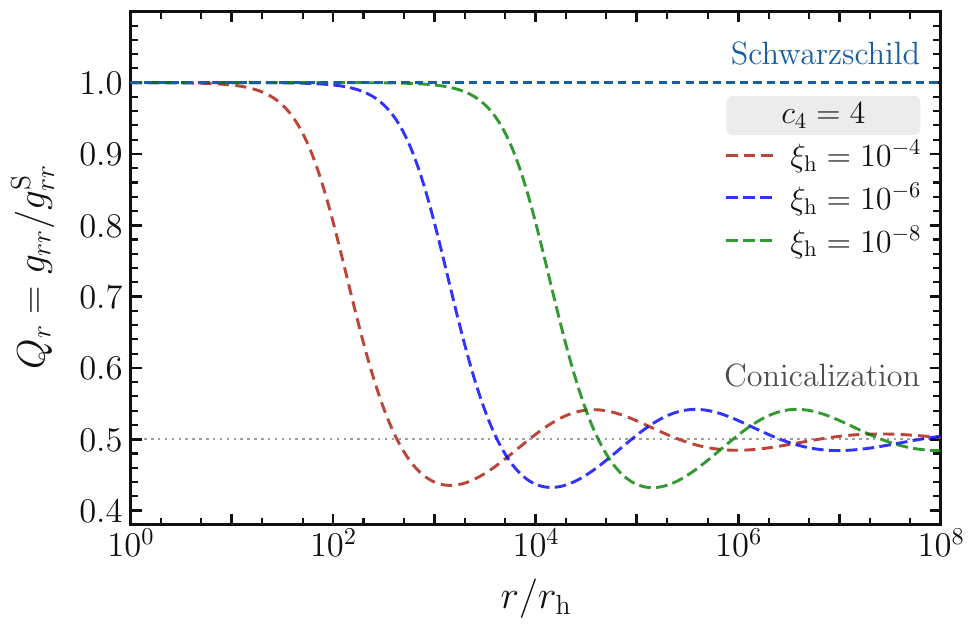}
  \caption{Ratio $Q_r=g_{rr}/g_{rr}^{\rm S}$ of the GCT radial metric component to the Schwarzschild value versus radius for $c_4=4$. Radius is measured in units of $\rh$. The red, blue, and green dashed profiles have $\xih=10^{-4}$, $10^{-6}$, and $10^{-8}$, respectively, with the red profile corresponding to the red dashed trajectory in Fig.~\ref{fig:manifolds}. The dark blue dashed line marks Schwarzschild, $Q_r=1$, and the gray dotted line marks the conical limit, $Q_r=1/2$. All profiles satisfy $Q_r\to1$ at the horizon. Weaker caloric fields shift the transition outward without changing the conical limit, which each profile approaches through damped radial oscillations. The label \textit{Conicalization} refers to this spatial transition within a static metric.}
  \label{fig:bh-profile}
\end{figure}

The numerical exteriors remain in the static region: $x_1$ stays positive without a further pole, and $x_2>0$ makes $e^\nu$ increase toward its asymptotic value. For a branch approaching $F_2$, the proper radial distance to infinity diverges. Moreover, $e^{\nu+\lambda}\to1-2/c_4>0$ implies that radial null geodesics require infinite affine parameter to reach $r\to\infty$. Together with the curvature decay in Eq.~(\ref{eq:cone-curvature}), these properties establish a regular outer end.

The two parameters play different physical roles. The model parameter $c_4$ fixes $F_2$ and the asymptotic angle, while the horizon amplitude $\xih$ selects the connecting trajectory. The exact solution enters an attracting neighborhood of $F_2$ at finite radius. By continuous dependence on regular horizon data, all sufficiently nearby solutions do so as well, establishing a continuous conical family around $\xih=1/c_4$ without an outer shooting condition. Numerical integration also gives the small-field members in Figs.~\ref{fig:manifolds} and \ref{fig:bh-profile} for the excess branch at $c_4=4$. For the deficit branch, integrations at $c_4=-4$ with $\xih=-10^{-4}$, $-10^{-6}$, and $-10^{-8}$ give an asymptotic solid angle of $8\pi/3$. At each $c_4$, the exact and small-field members share the same outer geometry despite their different intermediate profiles.

\subsubsection{Local Schwarzschild limit and conical transition}\label{sec:IV.D.2}
The independent caloric amplitude allows local deviations from Schwarzschild to become small without reducing the angular deformation at infinity. For either $c_4<0$ or $c_4>2$, we hold $\rh$ and $c_4$ fixed and take $\xih\to0$ within the allowed sign range. With a common local time normalization, the metric approaches Schwarzschild on every compact exterior interval. The expansion~(\ref{eq:horizon-series}) extends this convergence to the horizon in the regular ingoing chart. Nevertheless, every member approaching $F_2$ retains the solid angle $4\pi c_4/(c_4-2)$ in Eq.~(\ref{eq:solid-angle}), whereas Schwarzschild has $4\pi$. The two limits therefore do not commute:
\begin{subequations}\label{eq:nonuniform}
\begin{align}
  \lim_{\xih\to0}\lim_{r\to\infty}x_1 &= 1-\frac{2}{c_4}, \label{eq:nonuniform-conical}\\
  \lim_{r\to\infty}\lim_{\xih\to0}x_1 &= 1. \label{eq:nonuniform-schwarzschild}
\end{align}
\end{subequations}
The local Schwarzschild limit is thus nonuniform in radius. Figure~\ref{fig:manifolds} shows how this occurs in phase space: a trajectory with small nonzero $\xih$ first follows Schwarzschild toward $F_1$, then departs along its unstable direction and approaches $F_2$. Reducing $|\xih|$ delays the departure but leaves the final critical point unchanged.

Figure~\ref{fig:bh-profile} shows the corresponding radial metric profiles. To compare with Schwarzschild at the same areal radius and horizon radius, we define
\begin{equation}\label{eq:radial-metric-ratio}
  Q_r(r) \equiv \frac{g_{rr}}{g_{rr}^{\rm S}} = e^\lambda\left(1 - \frac{\rh}{r}\right).
\end{equation}

This ratio compares squared proper radial lengths for the same $\dx r$ and is independent of time normalization. Fixing $\rh$ compares equal horizon areas and does not assume equal asymptotic masses. Near the horizon, Eq.~(\ref{eq:horizon-f1}) gives $Q_r=1-\xih\varepsilon+\mathcal{O}(\varepsilon^2)$, so $Q_r\to1$ even though the horizon curvature depends on $\xih$. The red, blue, and green profiles in Fig.~\ref{fig:bh-profile} remain near $Q_r=1$ over increasingly large radial intervals. Each then falls below $Q_r=1/2$ and approaches this conical limit, corresponding to a solid angle of $8\pi$, through damped oscillations. The eigenvalues $-1/4\pm i\sqrt{15}/4$ at $F_2$ predict an asymptotic amplitude proportional to $(r/\rh)^{-1/4}$ and a period $8\pi/\sqrt{15}$ in $\ln(r/\rh)$, or approximately $2.82$ decades in radius. These are spatial oscillations of the static metric. They remain visible at the right edge of the plot, where the dotted line marks the limiting value rather than a finite integration endpoint.

For these profiles, we initialize $\mu(\eta)$ at $\eta=10^{-7}$ using the regular horizon series through order $\eta^2$. We integrate Eq.~(\ref{eq:master}) to $\eta=0.05$, then use its equivalent form in $\ln(r/\rh)$ to evaluate the exterior at fixed radii. Reducing the initial $\eta$ by a factor of ten and tightening the tolerances changes $Q_r$ by less than $10^{-8}$ over the plotted range. Independent integration of the original system~(\ref{eq:flow}) agrees to the same accuracy. Extending the integration beyond the plot confirms the predicted oscillation period and decay rate and supports the limits $x_2\to0$ and $x_3\to1/c_4$. Combined with the local sink analysis, these checks support a regular connection from $I_2$ to $F_2$ for each computed exterior.

To quantify the local departure from Schwarzschild, we introduce $\rho\equiv r/\rh$ and expand in $\xih$ at fixed $c_4$. Linearizing Eq.~(\ref{eq:normal}) and imposing the horizon value gives $\Fcal/x_1=\xih\rho^2+\mathcal{O}(\xih^2)$. The linearized field equation becomes $\dx[(\rho-1)x_3]/\dx\rho=\xih\rho^2+\mathcal{O}(\xih^2)$. Regularity at $\rho=1$ fixes the integration constant, yielding
\begin{subequations}\label{eq:linear-profile}
\begin{align}
  x_3 &= \frac{\xih}{3}(\rho^2 + \rho + 1) + \mathcal{O}(\xih^2), \label{eq:linear-field}\\
  \mu &= \frac{\xih}{3}(\rho^2 + 2\rho + 3) + \mathcal{O}(\xih^2). \label{eq:linear-mu}
\end{align}
The redshift gradient obtained from the same regular expansion is
\begin{align}\label{eq:linear-gradient}
  x_2 &= \frac{1}{\rho-1} + \frac{\xih\rho}{3(\rho-1)^2} \biggl(\frac{\rho^2}{2} + \rho \nonumber\\
    &\quad + \ln\rho + \frac{3}{\rho} - \frac{9}{2}\biggr) + \mathcal{O}(\xih^2).
\end{align}
\end{subequations}
The radial metric follows from $x_1=1+x_2-\mu$. The apparent poles in the correction to $x_2$ cancel, and its coefficient of $\xih$ tends to $1$ at the horizon. Integrating $\dx\nu/\dx\rho=x_2/\rho$ with the local normalization $f_{2{\rm h}}=1$ then gives
\begin{align}\label{eq:linear-lapse}
  \ln\frac{e^\nu}{1-\rho^{-1}} &= \xih\biggl[\frac{\rho}{6} + \frac{2}{3}\ln\rho \nonumber\\
    &\quad - \frac{\ln\rho}{3(\rho-1)} + \frac{1}{6}\biggr] + \mathcal{O}(\xih^2).
\end{align}
The expression in brackets vanishes at the horizon, as required by the normalization. These expansions have controlled remainders on compact exterior intervals. Their first-order coefficients are independent of $c_4$, which enters the local equations only through terms quadratic in the caloric field. To impose $e^\nu\to1$ at infinity, a constant time rescaling must be determined from the full exterior solution.

The growing term $x_3\propto\xih r^2/\rh^2$ in Eq.~(\ref{eq:linear-field}) reflects the unstable eigenvalue $2$ at $F_1$ and determines where the Schwarzschild expansion breaks down. Before this occurs, $\rho\gg1$ and $|\xih|\rho^2\ll1$ define a weak-field overlap region. At fixed $c_4$, Eqs.~(\ref{eq:linear-profile}) and (\ref{eq:linear-lapse}) give
\begin{subequations}\label{eq:weak-metric}
\begin{align}
  e^\nu &= 1 - \rho^{-1} + \frac{\xih\rho}{6} + \cdots, \label{eq:weak-temporal}\\
  e^\lambda &= 1 + \rho^{-1} - \frac{\xih\rho^2}{3} + \cdots. \label{eq:weak-radial}
\end{align}
\end{subequations}
The omitted growing terms are subleading in this overlap region. For large $|c_4|$, one also needs $|c_4\xih|\rho^2\ll1$ to suppress nonlinear field terms, so the expansion is not uniform in $c_4$. The meaning of agreement with Schwarzschild should be specified here, since weak-field tests probe both metric potentials~\cite{Will2014.LRR.17.4,Bertschinger2011.PhilTransRSocA.369.4947}. A correction can be small relative to the full metric coefficient but large relative to its Schwarzschild mass term. In the weak-field region, the fractional correction to the inverse-square force scales as $|\xih|\rho^2$, whereas the radial-metric correction relative to the mass term scales as $|\xih|\rho^3$. Together with the near-horizon expansion, these estimates give
\begin{equation}\label{eq:scale}
  \rh < r \ll \frac{\rh}{|\xih|^{1/3}},
\end{equation}
as the interval in which both metric corrections are small even relative to the Schwarzschild mass terms. At larger radii, the radial-metric correction can exceed the mass term while the full metric remains close to Schwarzschild. For fixed $c_4$ with an order-unity angular deformation, the spatial transition occurs on the scale $r_{\rm tr}\equiv\rh/\sqrt{|\xih|}$, up to a factor depending on $c_4$. The inverse-square force remains a good approximation throughout $\rh\ll r\ll r_{\rm tr}$, a wider interval than Eq.~(\ref{eq:scale}).

The transition requires the nonlinear equations. To isolate its radial scale, we take $\xih\to0$ at fixed $r/r_{\rm tr}>0$, $c_4$, and sign of $\xih$. The growing combination $\xih\rho^2=\operatorname{sgn}(\xih)(r/r_{\rm tr})^2$ remains finite, while matching to the inner solution gives $x_2/\sqrt{|\xih|}\approx r_{\rm tr}/r$ for small $r/r_{\rm tr}$. The equation $\dx\ln x_2/\dx\ln r=-x_1$ then preserves the scaling $x_2=\mathcal{O}(\sqrt{|\xih|})$ through the transition, where $x_1-1$ and $x_3$ are $\mathcal{O}(1)$. Thus $x_2\to0$, and the leading spatial profile is governed by Eq.~(\ref{eq:flow}) restricted to $x_2=0$. The same equation for $x_2$ determines the leading nonzero redshift gradient. The same spatial profile describes the limit $\rh\to0$ and $\xih\to0$ at fixed $r_{\rm tr}$, in which the local Schwarzschild mass vanishes. The resulting $x_2=0$ trajectory connects $F_1$ to $F_2$ and describes a horizonless spacetime with constant lapse. We call it the \textit{conically dressed Minkowski metric}: the caloric field connects an approximately Minkowskian core at $r\ll r_{\rm tr}$ to conical infinity.

The transition scalings explain how a finite spatial deformation can arise in an extremely weak gravitational field. Orthonormal components of the Riemann tensor are $r^{-2}$ times bounded functions of $x_i$, and are therefore $\mathcal{O}(|\xih|/\rh^2)$ at $r\sim r_{\rm tr}$. The proper acceleration of a static observer, $a_{\rm stat}=c^2x_2/(2r\sqrt{x_1})$, is $\mathcal{O}(c^2|\xih|/\rh)$ there. At fixed horizon scale, a weaker caloric field therefore moves the transition outward into a region of smaller curvature and acceleration. For the red, blue, and green profiles in Fig.~\ref{fig:bh-profile}, $r_{\rm tr}/\rh=10^2$, $10^3$, and $10^4$, respectively. Their first crossings of $Q_r=0.9$ occur at $r_{10}/\rh\approx61.9$, $618$, and $6.18\times10^3$, giving $r_{10}/r_{\rm tr}\approx0.618$ for this $c_4$ and threshold. Reducing $\xih$ by a factor of $100$ thus moves the transition outward by nearly a factor of $10$, and the rescaled profiles approach a common curve. Here $Q_r-1$ measures the fractional change in the full radial metric coefficient. Relative to the Schwarzschild mass term, the correction is instead $(g_{rr}-g_{rr}^{\rm S})/(g_{rr}^{\rm S}-1)=\rho(Q_r-1)$, whose smallness requires the stricter condition~(\ref{eq:scale}).

The conically dressed Minkowski metric and its caloric field are smooth at the center. On $x_2=0$, we normalize the constant lapse to $e^\nu=1$. Expanding Eq.~(\ref{eq:flow}) about $F_1$ gives $e^\lambda=1-\beta_1r^2+\mathcal{O}(r^4)$ and $\beta=\beta_1r+\mathcal{O}(r^3)$, where $[\beta_1]=\mathrm{length}^{-2}$. Both $(e^\lambda-1)/r^2$ and $\beta/r$ are analytic in $r^2$ on this branch. In local Cartesian coordinates $y^i$, with $r^2=\delta_{ij}y^iy^j$ and $y_i=\delta_{ij}y^j$, the spatial components become $g_{ij}=\delta_{ij}+[(e^\lambda-1)/r^2]y_iy_j$ and $S_i=(\beta/r)y_i$. These extend analytically through the center, with $S_i(0)=0$ and $\partial_jS_i(0)=\beta_1\delta_{ij}$.\footnote{For $\beta_1\ne0$, the radial component $\beta(|\boldsymbol{y}|)$ and the field magnitude $\sqrt{S_\mu S^\mu}\approx|\beta_1|\cdot|\boldsymbol{y}|$ are not differentiable at the origin. The vector field is nevertheless smooth, since its Cartesian components are analytic.} Along the $F_1\to F_2$ connection, finite $x_i$ and positive $x_1$ ensure smoothness at every positive radius, while the field decays as $1/r$ at infinity. The central curvature is finite, with $R(0)=-6\beta_1$ and $K(0)=12\beta_1^2$. Matching to Eq.~(\ref{eq:linear-field}) gives $|\beta_1|=1/(3r_{\rm tr}^2)$, so a large transition radius makes the central curvature arbitrarily small at fixed $c_4$.

\subsubsection{Comparison with other conical exteriors}
The independent caloric amplitude is the main distinction from familiar conical black-hole exteriors. Consider the Barriola-Vilenkin global-monopole metric, Eq.~(8) of Ref.~\cite{Barriola1989.PRL.63.341}, and the radial bumblebee metric, Eq.~(42) of Ref.~\cite{Casana2018.PRD.97.104001}. Both have an asymptotic solid angle of $4\pi/Q_r$, with $Q_r$ measured relative to Schwarzschild at the same horizon radius. For the monopole, this comparison uses the exterior approximation and requires the horizon to lie well outside the defect core. In both metrics, the fractional radial-metric deviation $Q_r-1$ is constant throughout this exterior description, including at the horizon, and is fixed by the asymptotic angle. Reducing the local deviation therefore reduces the angular deformation at infinity. In GCT, reducing $|\xih|$ suppresses the local deviation while $c_4$ and the asymptotic angle remain fixed.

The boundary conditions also differ. The prescribed flat-core monopole model of Ref.~\cite{BezerradeMello2007.CQG.24.2389} joins a regular inner region to a cone but has no black-hole horizon. For self-gravitating monopole black holes, an asymptotic scalar condition selects the scalar field value at the horizon~\cite{Watabe2004.JCAP.02.001}. The GCT construction requires no prescribed defect core or topological charge. Horizon regularity leaves $\xih$ free, and varying this amplitude within the constructed families changes the transition scale without assigning an outer field value. The near-Schwarzschild region can therefore extend to larger radii at the same asymptotic angle.

Figures~\ref{fig:manifolds} and \ref{fig:bh-profile} show the resulting geometry: an extended near-Schwarzschild region followed by a nonlinear transition to conical infinity. For a small caloric amplitude, the local expansion describes the region probed by bound orbits. We now use it to derive an observational test of the transition scale.

\section{Geodesics and observational tests}\label{sec:V}
The near-Schwarzschild region and the distant conical transition suggest complementary observational tests. Bound eccentric orbits probe local metric corrections through secular periapsis precession, while light rays can also probe the transition if their paths cross it. We first derive the leading orbital correction and the resulting constraint on the transition scale, then discuss light propagation in the full geometry.

Ordinary test matter follows metric geodesics in the present sector~\cite{Tian2026.PRD.114.044042}. For equatorial motion, the conserved quantities are the dimensionless specific energy $\mathcal{E}=e^\nu\dx t/\dx\tau$ and specific angular momentum $j=r^2\dx\varphi/\dx\tau$, where $\tau$ is proper time. The radial equation is
\begin{equation}\label{eq:radial-geodesic}
  e^{\nu+\lambda}\left(\frac{\dx r}{\dx\tau}\right)^2 + e^\nu\left(c^2 + \frac{j^2}{r^2}\right) = c^2\mathcal{E}^2.
\end{equation}
For a bound orbit with periapsis radius $r_{\rm p}$ and apoapsis radius $r_{\rm a}$, we choose the orientation so that $j>0$. The periapsis advance over one radial period is
\begin{equation}\label{eq:apsidal-angle}
  \Delta\varpi = 2\int_{r_{\rm p}}^{r_{\rm a}} \frac{je^{(\nu+\lambda)/2}\,\dx r}{r^2\sqrt{c^2\mathcal{E}^2-e^\nu(c^2+j^2/r^2)}} - 2\pi,
\end{equation}
where $\varpi$ is measured from a fixed direction in the orbital plane and positive $\Delta\varpi$ denotes prograde precession. This expression is invariant under a constant rescaling of time and can be evaluated directly with the full numerical metric.

To obtain an analytic prediction, we adopt the local time normalization $f_{2{\rm h}}=1$. The leading Newtonian potential is $-c^2\rh/(2r)$, whose coefficient $c^2\rh/2$ can be inferred from local orbital motion. This defines the relevant mass scale without referring to a global mass. For a reference Kepler ellipse with semimajor axis $a_{\rm orb}$ and eccentricity $0<e<1$, the mean motion is $n=\sqrt{c^2\rh/(2a_{\rm orb}^3)}$, and the turning radii are $r_{\rm p}=a_{\rm orb}(1-e)$ and $r_{\rm a}=a_{\rm orb}(1+e)$. The whole orbit remains in the weak-field perturbative region if $r_{\rm p}\gg\rh$ and $|\xih|(r_{\rm a}/\rh)^2\ll1$. For large $|c_4|$, one also needs $|c_4\xih|(r_{\rm a}/\rh)^2\ll1$ to suppress nonlinear field terms. Under these conditions, the caloric perturbation is small relative to Keplerian motion, although its periapsis correction can exceed the Schwarzschild contribution.

Using Eq.~(\ref{eq:weak-metric}) to expand the geodesic Lagrangian per unit mass, $\mathcal{L}=-c^2\dx\tau/\dx t$, gives the leading caloric perturbation
\begin{equation}\label{eq:caloric-lagrangian}
  \delta\mathcal{L} \approx -\frac{\xih}{6\rh^2}\left(r^2\dot{r}^2+\frac{c^2\rh r}{2}\right),
\end{equation}
where dots denote $\dx/\dx t$ and the ordinary Schwarzschild post-Newtonian terms are treated separately. The first and second terms arise from the radial and temporal metric corrections, respectively. For Keplerian motion at fixed eccentricity, the scaling $\dot r^2=\mathcal{O}(c^2\rh/r)$ places both terms at the same weak-field order. At first order, the Legendre transform gives $\delta\mathcal{H}\approx-\delta\mathcal{L}$ on the unperturbed orbit, with $\mathcal{H}$ also defined per unit mass. To extract the secular correction, we average over one Kepler period~\cite{Morbidelli2002.book}. Angle brackets denote the coordinate-time average $\langle X\rangle\equiv(n/2\pi)\int_0^{2\pi/n}X(t)\,\dx t$ of an orbital quantity $X(t)$. In terms of the eccentric anomaly $\psi$, $r=a_{\rm orb}(1-e\cos\psi)$ and $\dx t/\dx\psi=(1-e\cos\psi)/n$. Thus the anomaly integral carries the weight $1-e\cos\psi$. The averages $\langle r\rangle=a_{\rm orb}(1+e^2/2)$ and $\langle r^2\dot{r}^2\rangle=c^2\rh a_{\rm orb}e^2/4$ then give
\begin{equation}\label{eq:mean-hamiltonian}
  \langle\delta\mathcal{H}\rangle \approx \frac{c^2\xih a_{\rm orb}}{12\rh}(1+e^2).
\end{equation}
The two metric corrections contribute equally to the eccentricity-dependent part.

In the planar Delaunay variables for the reference Kepler orbit, $\varpi$ is canonically conjugate to $j=na_{\rm orb}^2\sqrt{1-e^2}$, and fixing the other action fixes $a_{\rm orb}$~\cite{Morbidelli2002.book}. Hamilton's equation and the chain rule therefore give the first-order secular correction
\begin{equation}\label{eq:secular-precession}
  \delta\dot{\varpi} \approx -\frac{\sqrt{1-e^2}}{na_{\rm orb}^2e}\left.\frac{\partial\langle\delta\mathcal{H}\rangle}{\partial e}\right|_{a_{\rm orb}}.
\end{equation}
The shift per orbit is $\delta\varpi=(2\pi/n)\delta\dot{\varpi}$. Including the leading Schwarzschild advance gives
\begin{subequations}\label{eq:precession}
\begin{align}
  \delta\varpi &\approx -\frac{2\pi\xih}{3}\left(\frac{a_{\rm orb}}{\rh}\right)^2\sqrt{1-e^2}, \label{eq:periapsis}\\
  \Delta\varpi &\approx \frac{3\pi\rh}{a_{\rm orb}(1-e^2)}+\delta\varpi. \label{eq:total-precession}
\end{align}
\end{subequations}
Here $\delta\varpi$ is the caloric contribution and $\Delta\varpi$ is the total advance, with mixed and higher-order terms omitted. The correction is retrograde for $\xih>0$ and prograde for $\xih<0$, corresponding to the excess and deficit branches, respectively. At this order it is independent of $c_4$. Since both metric potentials contribute equally, retaining only the temporal potential would miss half of the effect. The limit $e\to0$ describes slightly eccentric motion, as periapsis is undefined for a strictly circular orbit.

If an orbital fit bounds the additional precession rate by $|\delta\dot{\varpi}|\le\dot{\varpi}_{\max}$, the transition scale satisfies
\begin{equation}\label{eq:precession-sensitivity}
  r_{\rm tr} \gtrsim a_{\rm orb}\left[\frac{n\sqrt{1-e^2}}{3\dot{\varpi}_{\max}}\right]^{1/2}.
\end{equation}
The observable combination is $\xih/\rh^2$, whose magnitude is $r_{\rm tr}^{-2}$. Thus a precession bound constrains the transition scale but does not determine $c_4$ or the asymptotic solid angle at this order.\footnote{For a conditional estimate, suppose the solar exterior matches this horizon-selected branch with $\rh\approx2GM_\odot/c^2$. The INPOP15a Saturn interval $1.2\pm5.0$ milliarcseconds per century then corresponds to $|\xih|\sim10^{-26}$ or $r_{\rm tr}\sim1\,\mathrm{pc}$~\cite{Fienga2017.MG14.3694}. This interval uses a 35\% postfit-residual criterion, not a statistical confidence level. It only illustrates sensitivity, since neither a GCT ephemeris fit nor matching to a solar interior has been performed.} A quantitative constraint requires a joint fit of the local Kepler parameter, orbital initial conditions, and other correlated parameters, with a consistent treatment of orbital dynamics and light propagation~\cite{Fienga2024.LRR.27.1}. For black-hole orbits, rotation, extended matter, and other perturbers must also be included when relevant. Tighter precession bounds can therefore push the allowed transition to larger radii while remaining compatible with a finite angular deformation at infinity.

For light propagation, we define the ray parameter $b_\gamma\equiv r^2e^{-\nu}|\dx\varphi/\dx t|/c$, equal to $c$ times the ratio of the conserved angular-momentum magnitude to the conserved energy. This parameter has dimensions of length. Along a monotonic radial segment, a null trajectory obeys
\begin{equation}\label{eq:null-azimuth}
  \left|\frac{\dx\varphi}{\dx r}\right| = \frac{b_\gamma e^{(\nu+\lambda)/2}}{r^2\sqrt{1-e^\nu b_\gamma^2/r^2}}.
\end{equation}
For a static observer at radius $r$, the angle $\vartheta$ between the ray and the radial direction satisfies $\sin\vartheta=b_\gamma e^{\nu(r)/2}/r$. This combination is unchanged by a constant rescaling of time. A lensing calculation must follow the ray through the geometry along its path, including the conical transition if it is crossed. In the small-$|\xih|$ limit at fixed $c_4$ and sign of $\xih$, the additional azimuth relative to Schwarzschild over a fixed finite interval of $r/r_{\rm tr}>0$ is of order $b_\gamma/r_{\rm tr}$ for $b_\gamma\ll r_{\rm tr}$. This follows from $e^\nu=1+\mathcal{O}(\sqrt{|\xih|})$ and the possible finite departure of $e^\lambda$ from unity in that region (Sec.~\ref{sec:IV.D.2}). The observed deflection also depends on the source and observer positions and their local ray angles. Consequently, it cannot be inferred from the asymptotic solid angle alone.

\section{Conclusions and prospects}\label{sec:VI}
We identify spherical black holes beyond Schwarzschild in GCT by casting the fluid-free vacuum structure equations for a purely radial caloric field into a three-dimensional autonomous system. Regular conical infinity requires the model parameters to satisfy $c_6=2c_4$. On this parameter slice, the finite critical points $F_1$ and $F_2$ describe flat and conical infinity, respectively, with the latter allowed for $c_4<0$ or $c_4>2$. Regular horizons correspond to $I_2$ at phase-space infinity and restrict the admissible black-hole trajectories to its two-dimensional unstable manifold $\MH$. For $c_6=2c_4$, its intersection with the asymptotically flat manifold $\MAF(F_1)$ contains only the Schwarzschild trajectory. Figure~\ref{fig:manifolds} illustrates how other trajectories on $\MH$ connect $I_2$ to $F_2$, giving black holes with conical infinity. Taking the local Schwarzschild mass to zero at fixed transition scale gives the horizonless conically dressed Minkowski metric. Different metric classes occupy different \textit{volumes} in phase space: Minkowski occupies a point ($F_1$), the pure conical metric occupies a point ($F_2$), Schwarzschild occupies a curve ($I_2\to F_1$), conically dressed Minkowski occupies a curve ($F_1\to F_2$), and black holes with conical infinity occupy a two-dimensional surface ($\MH$, $I_2\to F_2$), while solutions with throats or physical singularities can occupy regions of nonzero three-dimensional volume.

Black holes with conical infinity and a large transition scale may be astrophysically relevant. For a sufficiently small caloric amplitude, these solutions can remain arbitrarily close to Schwarzschild from the near-horizon strong-field region to the conventional weak-field region (e.g., Solar System scales), yet retain an order-unity solid-angle deficit or excess at infinity. The red dashed trajectory with $\xih=10^{-4}$ in Fig.~\ref{fig:manifolds} shows this behavior in phase space: it closely follows Schwarzschild toward $F_1$ before departing toward $F_2$. The model parameter $c_4$ fixes the asymptotic solid angle $4\pi c_4/(c_4-2)$, whereas the caloric amplitude $\xih$ controls the transition scale $r_{\rm tr}=\rh/\sqrt{|\xih|}$. At fixed locally inferred mass, reducing $|\xih|$ moves the transition to larger radii. This freedom distinguishes the GCT solutions from the monopole exterior approximation of Ref.~\cite{Barriola1989.PRL.63.341} and the radial bumblebee metric of Ref.~\cite{Casana2018.PRD.97.104001}, whose fractional radial-metric deviations from Schwarzschild are fixed by their asymptotic angles and remain of the same order in the strong- and weak-field regions. Observationally, a bound on the leading GCT periapsis shift gives a lower limit on the transition scale through Eq.~(\ref{eq:precession-sensitivity}). This leading correction is independent of $c_4$ and therefore does not determine the asymptotic solid angle. Metrics of this kind illustrate how close agreement with general relativity on black-hole and Solar System scales can coexist with new physics in the extremely weak-field regime of local gravitational systems (rather than on cosmological scales), highlighting the need to develop observational probes of gravity in this regime~\cite{Jin2026.ApJL.1008.L41}.

In GCT, it remains unknown which types of black holes or other singular configurations can form through astrophysical processes. The phase space analyzed in this paper classifies static solutions, and a larger occupied \textit{volume} does not imply that a metric is favored by time evolution. Answering this question requires linear stability analyses of the static solutions~\cite{Regge1957.PhysRev.108.1063} and fully relativistic simulations of gravitational collapse~\cite{Choptuik1993.PRL.70.9}. As a first step, one can evolve regular caloric-field initial data without ordinary matter to test whether the solutions found here arise as late-time outcomes. If these evolutions produce black holes with conical infinity without prescribing a conical outer boundary, the process could be described as \textit{spontaneous conicalization} of black holes. This would be analogous to spontaneous black-hole scalarization, in which nonlinear simulations demonstrate the growth and saturation of scalar hair in suitable models~\cite{Ripley2020.CQG.37.155003,East2021.PRL.127.101102}. Static scalarized black-hole solutions~\cite{Doneva2018.PRL.120.131103,Silva2018.PRL.120.131104} can remain close to their general-relativistic counterparts in the weak-field region while differing substantially near the horizon. By contrast, for \textit{spontaneous conicalization}, the substantial departure from general relativity would occur in the extremely weak-field region.

\begin{acknowledgments}
This work was supported by the National Natural Science Foundation of China under Grant Nos.~12433001 and 12405050. OpenAI Codex assisted with language editing and mathematical checks.
\end{acknowledgments}

%

\end{document}